\documentclass{mem}
\usepackage{natbib}\usepackage{txfonts}\usepackage{balance}
\usepackage{graphicx}
\usepackage[latin1]{inputenc}
\usepackage[a4paper]{hyperref}
\idline{75}{282}
\begin{document}
\def\teff{$T\rm_{eff }$}
\def\kms{$\mathrm {km s}^{-1}$}
\def\kms{~km~s$^{-1}$}
\def\kmskpc{~km~s$^{-1}$~kpc$^{-1}$}
\def\cmmt{~cm$^{-3}$}
\def\cmmd{~cm$^{-2}$}
\def\smu{s$^{-1}$}
\def\um{$\mu$m}
\newcommand{\gsim}{\raisebox{-.4ex}{$\stackrel{>}{\scriptstyle \sim}$}}
\newcommand{\lsim}{\raisebox{-.4ex}{$\stackrel{<}{\scriptstyle \sim}$}}
\def\ls{\lsim}
\def\gs{\gsim}
\def\le{$\leq$}
\def\ge{$\geq$}
\def\deg{^\circ}
\def\lvd{$lv$-diagram}
\def\Msun{M$_\odot$}

\title{Gas dynamics in the Milky Way:}
\subtitle{the nuclear bar and the 3-kpc arms}

\author{N.J. \,Rodríguez-Fern\'andez\inst{1} 
}

  \offprints{N.J. Rodríguez-Fernández}

  \institute{
  IRAM, 300 rue de la Piscine, 38406 Saint Martin d'Heres, France 
    \email{rodriguez@iram.fr}
}

\authorrunning{Rodríguez-Fernández 
}

\titlerunning{Milky Way:  inner bar and  3-kpc arms}

\abstract{
We discuss the results of the first model of the gas dynamics in the Milky Way in the presence of two bars: the large scale primary bar or \emph{boxy} bulge and a secondary bar in the Galactic center region. We have obtained an accurate potential by modeling 2MASS star counts and we have used this potential to simulate the gas dynamics.
As a first approximation we have used one single pattern speed $\Omega_p$. The models with $\Omega_p=30-40$ \kmskpc \ and a primary bar orientation of $20º-35\deg$ reproduce successfully many characteristics of the observed longitude-velocity diagrams as the terminal velocity curve or the spiral arm tangent points. The Galactic Molecular Ring is not an actual ring but the inner part of the spiral arms, within corotation. The model reproduces \emph{quantitatively} the ``3-kpc arm" and the recently found far-side counterpart, which are the \emph{lateral arms} that contour the bar. 
In the Galactic center region, the model reproduces the 1-kpc HI ring and the Central Molecular Zone (CMZ), which is the gas response to the secondary bar.
In order to reproduce the observed parallelogram shape of the CO longitude velocity diagram of the CMZ, the secondary bar should be oriented by and angle of $60º-70º$ with respect to the Sun-GC line. The mass of the secondary bar amounts to $(2-5.5)\,10^9$~\Msun, which is 10-25 \% of the mass of the primary bar.

}

\maketitle{}

\section{Introduction}

The existence of a bar or boxy bulge in the Milky Way is now well established. However,
the structure and the dynamics of the Galaxy still present many unknowns as
the exact orientation of this bar and its  pattern speed.
Recently, 
 \cite{Alard01} and \cite{Nishiyama05} have found evidence of a secondary bar in the inner 4 degrees of the Galaxy. The exact nature of this component and the effect on the gas dynamics in the Galactic center  and the ``Central Molecular Zone" (CMZ)
remain to be studied in detail.
Other open questions regarding the gas dynamics,  are  the exact nature of some intense features of the CO longitude-velocity diagram (hereafter \lvd)  as the ``Galactic Molecular Ring" (GMR), the ``Connecting Arm" and the ``3-kpc arm".

\begin{figure*}[t!]
\centerline{\includegraphics[angle=-90,width=8cm]{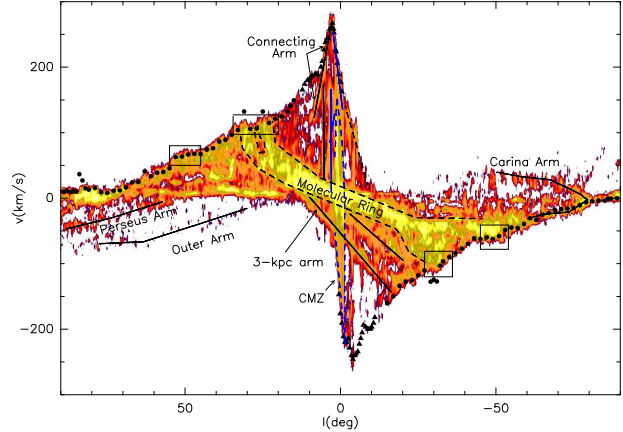}}
\caption{\footnotesize
Longitude-velocity ($lv$) diagram of the CO(1-0) emission \citep{Dame01}. The solid lines trace the position of some remarkable features such as the locus of the spiral arms, the 3-kpc arm and the Connecting Arm. {  The black dashed lines indicate the contour of the Galactic Molecular Ring. The solid circles and the triangles are the terminal velocities measurements using CO and HI, respectively. 
The boxes mark the position of the Sagittarius, Scutum, Norma and Centaurus tangent points. The lines concerning the Nuclear Disk or Central Molecular Zone are blue but they are better shown in the next figure. } All these lines are used to compare with the simulations results in Fig. 3.
}
\label{fig:dame}
\end{figure*}

In this paper we will focus our attention to the inner Galaxy and in particular to the ``3-kpc arm" and its recently found far-side counterpart \citep{Dame08} and to the CMZ and its interplay with the secondary or nuclear stellar bar. The framework of our analysis will be the recent  simulations by \cite{Rodriguez08} (hereafter RFC08), who have modeled the gas dynamics in the presence of two bars using a realistic potential derived with 2MASS star counts.

\begin{figure}[t!]
\resizebox{\hsize}{!}{\includegraphics*[bb=49 175 519 461, clip=true,angle=0]{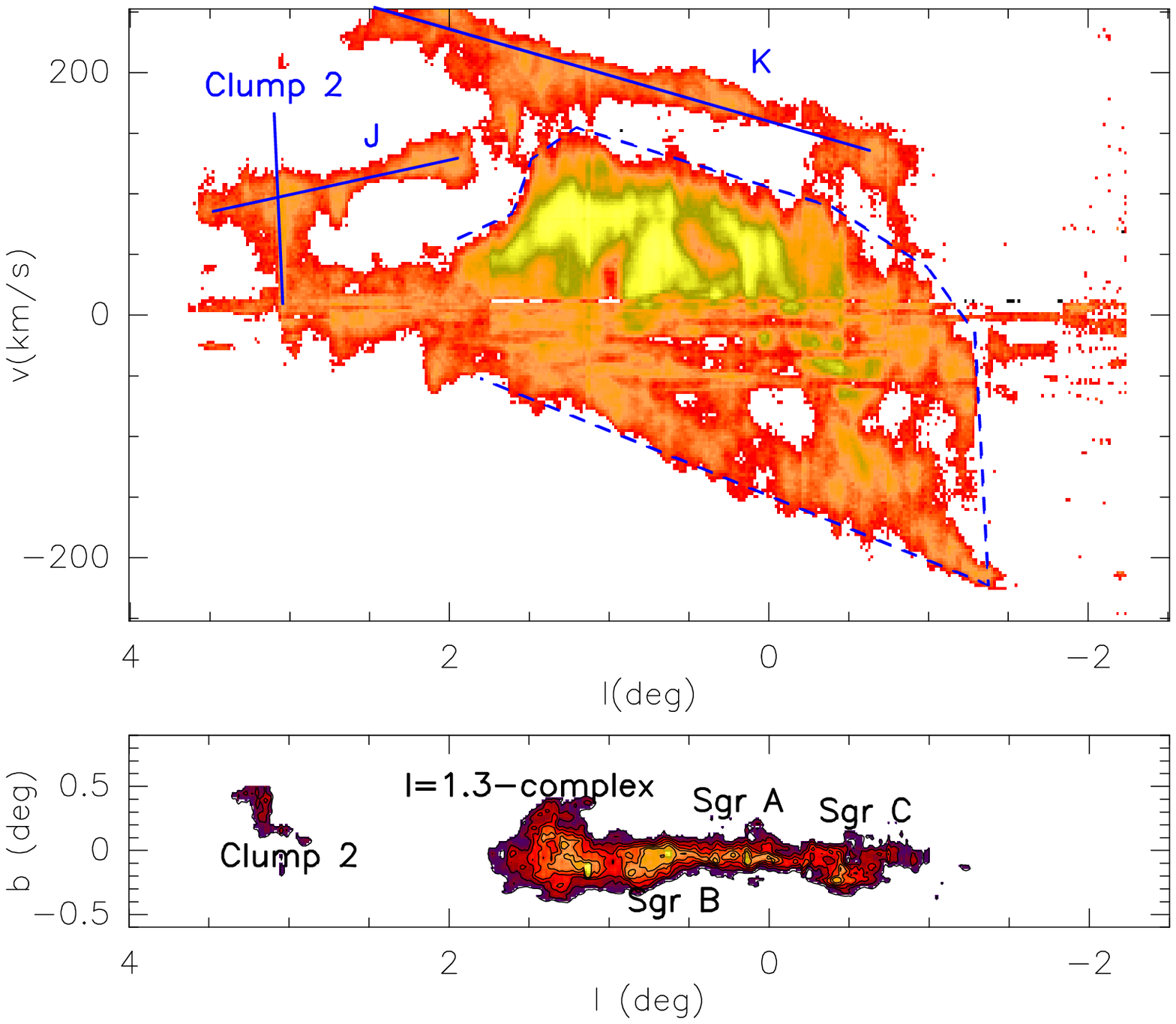}}
\caption{\footnotesize
Longitude-velocity ($lv$) diagram of the CO(1-0) emission in the Central Molecular Zone using data of \cite{Bally87} and integrating all negative Galactic latitudes. The solid lines trace the position of some remarkable features such as  Clump 2, or structures  K and J of \citep{Rodriguez-Fernandez06}. The dashed line indicates the contour of the very crowded central region.
}
\label{fig:bally}
\end{figure}
%

\section{The pattern speed and orientation of the primary bar}
The RFC08 models with $\Omega_p=30-40$ \kmskpc \ and a bar orientation of $20º-35\deg$ reproduce successfully many characteristics of the observed \lvd \ as the terminal velocity curve or the spiral arm tangent points. The Galactic Molecular Ring is also reproduced. However, it is not an actual ring but the inner part of the spiral arms, within corotation. The 3-kpc arm and its far-side counter part, the HI 1-kpc ring and the CMZ are also reproduced (see below). 

The inclination of the bar derived by RFC08 is in good agreement with previous determinations. Regarding the bar pattern speed, the RFC08 results  are
in good agreement with those of \cite{Weiner99}, $\Omega_p=42$ \kmskpc, or \cite{Fux99}. In his simulations  the pattern speed evolves from 50 to 30 \kmskpc .
\cite{Ibata95} have obtained  $\Omega_p=22-28.4$ \kmskpc \  by fitting K stars data with non-axisymmetric models.
In contrast, \cite{Englmaier99}  suggest $\Omega_p=50-60$ \kmskpc.
At the time being, it is fair to say that the pattern speed of the Milky Way is in the range $30-60$ \kms.

\section{The 3-kpc arm and its far-side counterpart}
The 3-kpc arm is clearly seen in the $lv$-diagram of Fig. \ref{fig:dame} with a velocity of -53 km\,s$^{-1}$ at $l=0\deg$.
 The simulations of \cite{Fux99} suggest that it is a lateral arm that surrounds the bar while, in the simulations by \cite{Englmaier99}, it would be a small arm arising from the extremity of the bar.
\cite{Habing06} proposed a different explanation based on the finding of old stars associated with the 3-kpc arm. 
They interpret the fact that both old stars and gas could follow the same trajectories as the probe
as that the arm would be the locus of closed orbits and not a spiral density wave maximum. They propose that the 3-kpc arm has its origin near one of the two points where the bar meets its corotation radius and that the arm can be a channel to transport gas from the corotation to the GC,  fueling the star formation. 

\begin{figure*}[t!]
\begin{center}
{\includegraphics[angle=-90, width=11cm]{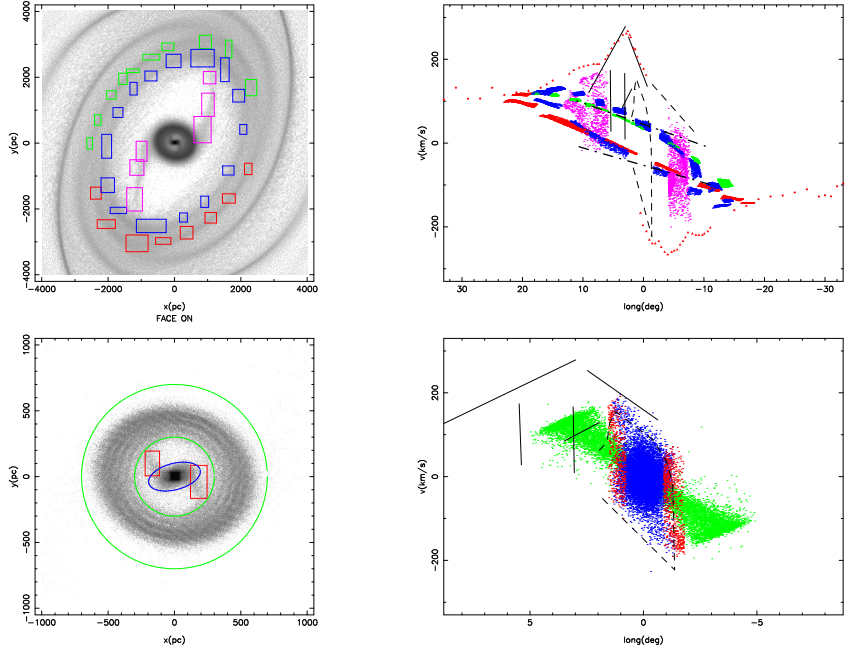}}
{\includegraphics[clip=true,angle=-90,width=11cm]{rodriguez3b.eps}}
\end{center}
\caption{\footnotesize
This figure shows the face on view of the inner Galaxy for the reference model by RFC08 (pattern speed $\Omega_p=30$~\kmskpc, orientation of the primary and secondary bar $\alpha^B=20\deg$ and $\alpha^b=75\deg$, respectively). The Sun is located at  $(0,-8000)$. The upper left panel shows the inner $8\times8$ kpc$^2$ and the lower left panel a zoom of the $2\times2$ kpc$^2$.
Several regions have been selected in the face on views to identify the locus of the different structures in the \lvd \ of the right panels  (using the same color code in the face on view and the \lvd).
  Solid and dashed lines, circles and triangles are the same as in the previous figures. The dot-dashed lines in the upper \lvd \ are the fits to the near and far 3-kpc arms as given by \cite{Dame08}.
}
\label{fig:simu}
\end{figure*}

The middle panels of Fig. \ref{fig:simu} show the locus of the lateral arms in the simulations by RFC08, which reproduce quite well the structure of the 3-kpc arm.
Therefore, the most likely explanation for the 3-kpc arm is a lateral arm. 
This is in agreement with \cite{Fux99}.
However, the 
\cite{Fux99} simulated Galaxy is rather asymmetrical and he identified  the ``135 \kms \ arm'' as the far-side counterpart of the ``3-kpc arm". While in our models the far side 3-kpc is symmetrical to the 3-kpc arm, in agreement with the recent finding by \cite{Dame08} of an arm behind the Galactic center, that is approximately simmetrical to the 3-kpc arm and that they have also identified as the far-side counterpart of the 3-kpc arm.

\section{The secondary bar and the Central Molecular Zone}

The Central Molecular Zone (hereafter CMZ) refers to the gas accumulation in the inner hundreds of parsec of the Galaxy. 
The CMZ extends from approximately  $-1.5\deg$ to $2\deg$. The spatial distribution is not symmetric due to the prominent cloud complex located at $l=1.3\deg$  (hereafter {\it l=1.3$\deg$-complex}).
Figure \ref{fig:bally} shows the \lvd \ of the inner degrees of the Galaxy.
The kinematics of this region are very complex with a high fraction of the gas exhibiting non-circular velocities. The contour of the inner CMZ in the \lvd \ is shown with a dashed line in Fig. \ref{fig:bally}.

In the central kpc of the RFC08 Galaxy there is a ring and inside this ring there is a small bar-like structure that is connected to the ring by two small spiral arms.
The  ring size is in excellent agreement with the observed HI ring.
The elongated structure found in the central hundreds of parsecs corresponds to the gas response to the nuclear bar and it resembles the observed CMZ.
The   \lvd\  of these structure reproduces the parallelogram of Fig. \ref{fig:bally} for a secondary bar orientation of $\sim 75º$. This value is in excellent agreement with the face on view of the CMZ inferred from CO and OH data by \cite{Sawada04}.
In the RFC08 models, the velocity dispersion of the CMZ is well reproduced for a mass of the nuclear bar of $2.0-5.5\,10^{9}$ M$_\odot$

The RFC08 simulations are the very first attempt to model the Milky Way with two nested bars and have been performed assuming a common pattern speed for both bars. Therefore, they could not study in detail the  coupling and the evolution of the nuclear bar.
Nevertheless,  these simulations explain many characteristics of the Galaxy at scales from the disk to the nuclear region, some for the first time, such as the parallelogram of the CMZ.
Therefore, at present, the observational data are \textit{compatible} with a scenario of coupled bars rotating with the same speed.

RFC08 have also proposed an scenario to explain the observed lopsidedness of the CMZ: it can be due to material falling into the CMZ from the HI ring through only one of the inner spiral arms.
Fig. \ref{fig:simu} shows that the clouds in the arm seen at negative longitudes will be  outside the observed parallelogram of the \lvd \ while the clouds in the arm seen at
positive longitudes  ``fill" the observed parallelogram. In this context, strong shocks are expected in the interaction region.
Indeed the 
{\it l=1.3$\deg$-complex}  shows the signature of strong shocks \citep[][RFC08]{Rodriguez-Fernandez06}.

\bibliographystyle{aa}

\begin{thebibliography}{}

\bibitem[Alard (2001)]{Alard01}
{Alard}, C. 2001, \aap, 379, L44

\bibitem[{{Bally} {et~al.}(1987){Bally}, {Stark}, {Wilson}, \&
  {Henkel}}]{Bally87}
{Bally}, J., {Stark}, A.~A., {Wilson}, R.~W., \& {Henkel}, C. 1987, \apjs, 65,
  13


\bibitem[Dame \& Thaddeus(2008)]{Dame08} Dame, T.~M., \& Thaddeus, P.\ 2008, \apjl, 683, L143 

\bibitem[{{Dame} {et~al.}(2001){Dame}, {Hartmann}, \& {Thaddeus}}]{Dame01}
{Dame}, T.~M., {Hartmann}, D., \& {Thaddeus}, P. 2001, \apj, 547, 792


\bibitem[{{Englmaier} \& {Gerhard}(1999)}]{Englmaier99}
{Englmaier}, P. \& {Gerhard}, O. 1999, \mnras, 304, 512

\bibitem[{{Fux}(1999)}]{Fux99}
{Fux}, R. 1999, \aap, 345, 787

\bibitem[{{Habing} {et~al.}(2006){Habing}, {Sevenster}, {Messineo}, {van de
  Ven}, \& {Kuijken}}]{Habing06}
{Habing}, H.~J., {Sevenster}, M.~N., {Messineo}, M., {van de Ven}, G., \&
  {Kuijken}, K. 2006, \aap, 458, 151

\bibitem[Ibata 
\& Gilmore(1995)]{Ibata95} Ibata, R.~A., \& Gilmore, G.~F.\ 1995, \mnras, 275, 605 



\bibitem[{{Nishiyama} {et~al.}(2005){Nishiyama}, {Nagata}, {Baba}, {Haba},
  {Kadowaki}, {Kato}, {Kurita}, {Nagashima}, {Nagayama}, {Murai}, {Nakajima},
  {Tamura}, {Nakaya}, {Sugitani}, {Naoi}, {Matsunaga}, {Tanab{\'e}},
  {Kusakabe}, \& {Sato}}]{Nishiyama05}
{Nishiyama}, S., {Nagata}, T., {Baba}, D., {et~al.} 2005, \apjl, 621, L105

\bibitem[{{Rodriguez-Fernandez} {et~al.}(2006){Rodriguez-Fernandez}, {Combes},
  {Martin-Pintado}, {Wilson}, \& {Apponi}}]{Rodriguez-Fernandez06}
{Rodriguez-Fernandez}, N.~J., {Combes}, F., {Martin-Pintado}, J., {Wilson},
  T.~L., \& {Apponi}, A. 2006, \aap, 455, 963

\bibitem[Rodriguez-Fernandez \& Combes, (2008)]{Rodriguez08} Rodriguez-Fernandez, N.~J., \& Combes, F.\ 2008, \aap, 489, 115 


\bibitem[{{Sawada} {et~al.}(2004){Sawada}, {Hasegawa}, {Handa}, \&
  {Cohen}}]{Sawada04}
{Sawada}, T., {Hasegawa}, T., {Handa}, T., \& {Cohen}, R.~J. 2004, \mnras, 349,
  1167
  
\bibitem[{{Weiner} \& {Sellwood}(1999)}]{Weiner99}
{Weiner}, B.~J. \& {Sellwood}, J.~A. 1999, \apj, 524, 112


\end{thebibliography}

\end{document}